\documentclass[page-classic,linenumbers]{nrc2} 

\usepackage{graphicx}
\usepackage{amsmath}
\usepackage{amssymb}
\usepackage{xcolor}
\usepackage{bm}       
\usepackage{flafter}
\usepackage{float}

\def\im{ \,\mathrm{Im}\,}
\def\re{ \,\mathrm{Re}\,}

\begin{document}

\title{The effect of strong electron-rattling phonon coupling on some superconducting properties}

\author{Samin Tajik}
\address[brock]{Department of Physics, Brock University, St. Catharines, 
Ontario L2S 3A1, Canada}
\author{Bo\v zidar Mitrovi\'c}
\address[brock]
\author{Frank Marsiglio}
\address{Department of Physics, University of Alberta, Edmonton, Alberta T6G 2E1, Canada} 
\shortauthor{S. Tajik {\it et al.}}



\begin{abstract}
Using the Eliashberg theory of superconductivity we have examined several properties of a model in which electrons are coupled only 
to rattling phonon modes represented by a sharp peak in the electron-phonon coupling function. Our choice of parameters was guided by
experiments on $\beta$-pyrochlore oxide superconductor KOs$_{2}$Os$_{6}$. We have calculated the temperature dependence of the 
superconducting gap edge, the quasiparticle decay rate, the NMR relaxation rate assuming that the coupling between the nuclear spins and the 
conduction electrons is via a contact hyperfine interaction which would be appropriate for the O-site in KOs$_{2}$Os$_{6}$, and the 
microwave conductivity. We examined the limit of very strong coupling by
considering three values of the electron-phonon coupling parameter $\lambda=$ 2.38, 3, and 5 and {\em did not} assume that the rattler
frequency $\Omega_{0}$ is temperature dependent in the superconducting state. We obtained a very unusual temperature dependence of
the superconducting gap edge $\Delta(T)$, very much like the one extracted from photoemission experiments on KOs$_2$O$_6$.

\end{abstract}

\maketitle*

\section{Introduction}

The $\beta$-pyrochlore osmium oxide superconductors AOs$_{2}$Os$_{6}$ 
(A=Cs, Rb, K) have been extensively studied experimentally and theoretically since  
the discovery of superconductivity in these compounds in 2004 (for reviews  
see \cite{Nagao,Hiroi}). The key structural feature of this class of 
compounds is that alkali ion A is located in an oversized cage composed 
of OsO$_{6}$ octahedra and moves in a flat anharmonic potential well 
\cite{Kunes}. This motion results in almost localized anharmonic 
modes at low energies -- so-called rattling modes. The rattling modes
are responsible for many of the observed physical properties of 
$\beta$-pyrochlores, such as the nuclear magnetic resonance (NMR) 
relaxation rate 1/($T_{1}T)$ of potassium nucleus in KOs$_{2}$Os$_{6}$ 
which is dominated by coupling of the electric field gradient to the 
nuclear quadrupole moment \cite{Yoshida,Dahm}. As the size of the alkali 
ion is reduced from Cs to K the potential in which the ion moves 
becomes flatter and wider \cite{Kunes} which results in an increased 
anharmonicity. At the same time the superconducting transition 
temperature $T_{c}$ increases from 3.25 K for A=Cs to 9.60 K for A=K 
leading to a conclusion that rattling modes play an important role in 
superconductivity of these compounds, in particular because the 
electronic structures of these compounds and the electronic 
density of states at the Fermi level are almost identical to each 
other (see \cite{Hiroi}, section 4.1, and the references therein). 

The most direct evidence that the low frequency phonon modes play an important 
role in superconductivity of KOs$_{2}$Os$_{6}$ comes from the photoemission 
spectroscopy measurements by Shimojima et al. \cite{Shimojima} of the 
electronic density of states (DOS) in superconducting state $N_{s}(\omega)$. The observed peak 
at 3.7 meV followed by a dip at 4.5 meV (see the inset in Fig. 1(a) in \cite{Shimojima})  
are directly related to a sharp peak in the electron-phonon coupling function 
$\alpha^{2}(\Omega)F(\Omega)$ at energy $\Omega_{0}$ equal to 3.7 meV  minus the gap 
edge $\Delta(T)$ = 1.63 meV at $T$ = 4.7 K. For excitation energy $\omega\leq\Omega_{0}
+\Delta(T)$ the electron-phonon interaction is attractive and the real part of the gap 
function $\Delta_{1}(\omega)=\re \Delta(\omega)$ is positive and attains maximum near 
$\omega=\Omega_{0}+\Delta(T)$ caused by resonant exchange of (virtual) phonons (see Fig. 34 in 
\cite{Scalapino}). At the same time a quasiparticle added at energy $\omega$ near 
$\Omega_{0}+\Delta(T)$ can decay down to the gap edge $\Delta(T)$ by emitting 
a {\em real} phonon of energy $\Omega_{0}$ leading to an increase of the quasiparticle damping rate which is given 
by the modulus of the imaginary part of the gap $\Delta_{2}(\omega)=\im \Delta(\omega)$. As a  
result, the electronic density of states in the superconducting state

\begin{equation}
\label{DOS} \re\frac{\omega}{\sqrt{\omega^{2}-\Delta^{2}(\omega)}}\approx
            1+\frac{\Delta_{1}^{2}(\omega)}{2\omega^{2}}-\frac{\Delta_{2}^{2}(\omega)}{2\omega^{2}}\>
\end{equation}
shows a peak near $\Omega_{0}+\Delta(T)$ followed by a dip (see Fig. 38 in \cite{Scalapino}). The 
energy $\Omega_{0}\approx$ 2.1 meV of the rattling mode obtained from photoemission spectroscopy 
measurements in the superconducting state of KOs$_{2}$Os$_{6}$ \cite{Shimojima} is very similar to 
the value of 2.3 meV obtained by Dahm and Ueda \cite{Dahm} in their fits of the NMR relaxation rate 
of the potassium nucleus in this compound.

Several experiments point independently to a very strong electron-phonon coupling in $\beta$-pyrochlores, in 
particular in the case of KOs$_{2}$Os$_{6}$. The ratio of the measured Sommerfeld coefficient $\gamma$ of 
the normal state electronic specific heat to the value found in band structure calculations $\gamma_{\mbox{band}}$ is
$\gamma/\gamma_{\mbox{band}}$ = 7.3 \cite{Hiroi}. Since enhancement of magnetic susceptibility in $\beta$-pyrochlores
is nearly absent \cite{ Hiroi}, a large electron mass enhancement implied by a large value $\gamma/\gamma_{\mbox{band}}$ 
is caused by the electron-phonon interaction, and in that case theory \cite{Grimvall} gives $\gamma/\gamma_{\mbox{band}}$  
= 1$+\lambda$, where 
\begin{equation}
\label{lambda} \lambda=2\int_{0}^{+\infty}d\Omega\frac{\alpha^{2}(\Omega)F(\Omega)}{\Omega}\>.
\end{equation}
Thus, for KOs$_{2}$Os$_{6}$ $\lambda$ = 6.3 which is the largest value for any known electron-phonon 
superconductor. This value is essentially identical with the one obtained from de Haas-van Alphen oscillation 
measurements and a band structure calculation \cite{Terashima} (see Fig. 3 in \cite{Terashima}). In \cite{Terashima} 
it was found that orbit-resolved mass enhancements are homogeneous with the $\lambda$-values concentrated in the range 5-8.

The photoemission spectroscopy measurements \cite{Shimojima} obtained temperature dependent superconducting gap edge 
$\Delta(T)$ and found 2$\Delta(0)/k_{B}T_{c}\geq$ 4.56 which is indicative of a very strong electron-phonon 
coupling in KOs$_{2}$Os$_{6}$ \cite{mitrovic84,CMM} (the weak-coupling BCS value is 3.53). The measured ratio of 
the jump $\Delta C$ in the specific heat at $T_{c}$ to the normal state electronic specific heat $\gamma T_{c}$ is 
$\Delta C/\gamma T_{c}$ = 2.87 \cite{Hiroi07}, which is much larger than the weak-coupling BCS value of 1.43 and indicates 
strong electron-phonon coupling \cite{Marsiglio}. Similarly, the measured ratio $\gamma T_{c}^{2}/H_{c}^{2}(0)$, where 
$H_{c}(0)$ is the thermodynamic critical field at zero temperature, is 0.128 \cite{Hiroi} (the BCS value is 0.168), which 
is one among the smallest values for strong coupling superconductors \cite{Marsiglio} (Pb-Bi alloy shown in Fig.4 of 
reference \cite{Marsiglio} should be Pb$_{.65}$Bi$_{.35}$). The thermodynamic critical field deviation function 
$D(t)=H_{c}(t)/H_{c}(0)-(1-t^2)$, $t=T/T_{c}$, was measured in a limited temperature range \cite{Hiroi07} above 8.5 K 
\cite{Nagao} (see Fig. 12 in \cite{Nagao}) and the values were positive, slightly below those for Pb, suggesting 
strong electron-phonon coupling in KOs$_{2}$Os$_{6}$. In $\beta$-pyrochlores the alkali atom donates one electron to 
the cage making it metallic and the measurements of the NMR relaxation rate on O-sites \cite{Yoshida} probed the spin dynamics 
of conduction electrons. These measurements on KOs$_{2}$Os$_{6}$ found that the Hebel-Slichter peak below $T_{c}$ is strongly suppressed which
is consistent with very strong conduction electron-phonon coupling in this compound \cite{Akis,Allen}.  

In this work we consider a model system in which superconductivity arises exclusively from the conduction electron coupling to a rattling 
phonon modeled by a single sharp peak in $\alpha^{2}(\Omega)F(\Omega)$ at rattling frequency $\Omega_{0}$ \cite{Mahan}. 
We do not aim to reproduce the experimental results obtained for KOs$_{2}$Os$_{6}$, although our choice of parameters, such as $\Omega_{0}$ and 
$\lambda$, is motivated by what was obtained for this compound.  
One earlier study \cite{Marsiglio91} examined the behavior of the gap function and density of states in the {\it extreme} 
strong coupling limit. Certainly the applicability of Eliashberg theory is questionable for the extreme strong coupling parameters explored 
in \cite{Marsiglio91} and here. Nonetheless, we further and explore 
the temperature dependence of the gap edge $\Delta(T)$, 
and various dynamic and thermodynamic properties in such an extreme case. Before one tries to model KOs$_{2}$Os$_{6}$ by making a more 
realistic choice for $\alpha^{2}(\Omega)F(\Omega)$ which includes electron coupling to phonon modes other than the rattling one, it 
is important to delineate the effects of electron-rattler coupling. 

\section{Model and theoretical background}

We model the $\alpha^{2}(\Omega)F(\Omega)$ of electrons coupled to the rattling phonon with a cut-off Lorentzian
centered at energy $\Omega_{0}$ and having the half-width $\varepsilon$
\begin{equation}
\label{a2F}
\alpha^{2}(\Omega)F(\Omega)=\frac{g\varepsilon}{\pi}\left[\frac{1}{(\Omega-\Omega_{0})^{2}+
\varepsilon^{2}}-\frac{1}{\Omega_{c}^{2}+\varepsilon^{2}}\right]\>,
\end{equation}
for $\vert\Omega-\Omega_{0}\vert\leq\Omega_{c}$ and zero otherwise. We assume that at superconducting 
temperatures $\Omega_{0}$ and $\varepsilon$ are temperature-independent. Motivated by experiments on 
KOs$_{2}$Os$_{6}$ we take $\Omega_{0}=$ 2.2 meV and choose $\varepsilon=$ 0.01 meV and $\Omega_{c}=$ 2.1 meV.
The strength $g$ in (\ref{a2F}) was chosen to obtain a particular value for $\lambda$ given by (\ref{lambda}), 
and we considered $\lambda=$ 2.38, 3, and 5.

First, Eliashberg equations on the imaginary axis \cite{AM,Carbotte} were solved for the superconducting transition 
temperature $T_{c}$. At $T=T_{c}$ these equations can be cast into a Hermitian eigenvalue problem
\begin{equation}
\label{TC}
\bar{\phi}(n)=e(T)\sum_{\vert\omega_{m}\vert\leq\omega_{c}}\pi T\frac{\lambda(\omega_{n}-\omega_{m})-\mu^{*}
(\omega_{c})}{\sqrt{\vert\omega_{n}\vert Z(n)
\vert\omega_{m}\vert Z(m)}}\bar{\phi}(m)\>,
\end{equation}
and $T_{c}$ is the highest temperature $T$ at which the largest eigenvalue $e(T)$ is equal to 1. In (\ref{TC}) 
$\omega_{c}$ is a cutoff on Matsubara frequencies $\omega_{n}=\pi T(2n-1)$ and $\mu^{*}(\omega_{c})$ is 
the Coulomb pseudopotential for that cutoff. We took $\omega_{c}=$ 30 meV ($\gg \Omega_{0}$) and $\mu^{*}(\omega_{c})=$ 0.1 
for all values of $\lambda$. Furthermore, in (\ref{TC}) $\bar{\phi}(n)=\phi(n)/\sqrt{\vert\omega_{n}\vert Z(n)}$, 
where $\phi(n)$ is the pairing self-energy at $i\omega_{n}$, and 
\begin{equation}
\label{Z}
Z(n)=1+\frac{\pi T}{\omega_{n}}\sum_{m=-\infty}^{+\infty}\lambda(\omega_{n}-\omega_{m})\frac{\omega_{m}}
{\vert\omega_{m}\vert}
\end{equation}
is the renormalization function at $i\omega_{n}$. In (\ref{TC}) and (\ref{Z})   
\begin{equation}
\label{lambdanm}
\lambda(\omega_{n}-\omega_{m})=\int_{0}^{\infty}d\Omega\alpha^{2}(\Omega)F(\Omega)
\frac{2\Omega}{\Omega^2+(\omega_{n}-\omega_{m})^{2}}
\end{equation}
is the electron-phonon kernel at temperature $T$.

Next, the gap function $\Delta(\omega)$ and the renormalization function $Z(\omega)$ are 
obtained by solving the Eliashberg equations at finite temperature on the real axis 
\begin{eqnarray}
\label{Eli1}
\phi(\omega) & = &\int\limits_0^{\omega_c}d\omega'\re[M(\omega')]\left[
f(-\omega')K^{+}(\omega,\omega')\right. \\ 
                        & - &f(\omega')K^{+}(\omega,-\omega')-\mu^{*}(\omega_c)\tanh\frac{\omega'}{2T} \nonumber\\
	                & + &\left. \bar{K}^{+}(\omega,\omega')-\bar{K}^{+}(\omega,-\omega')\right] \>, \nonumber 
\end{eqnarray}
\begin{eqnarray}
\label{Eli2}
Z(\omega))& = &1-\frac{1}{\omega}\int\limits_0^{+\infty}d\omega'\re[N(\omega')]\left[ f(-\omega')\right. \\
          &\times&  K^{-}(\omega,\omega')-f(\omega')K^{-}(\omega,-\omega') \nonumber\\
          & + &\left. \bar{K}^{-}(\omega,\omega')-\bar{K}^{-}(\omega,-\omega')\right] \nonumber\>,
\end{eqnarray}
where $\phi(\omega)=\Delta(\omega)Z(\omega)$ is the pairing self-energy, 
\begin{equation}
\label{ADOS}
M(\omega)=\frac{\Delta(\omega)}{\sqrt{\omega^{2}-\Delta^{2}(\omega)}}\>,
\end{equation}
\begin{equation}
	N(\omega)=\frac{\omega}{\sqrt{\omega^{2}-\Delta^{2}(\omega)}}\>,
\end{equation}
and $f$ is the Fermi function.
The electron-phonon coupling function $\alpha^{2}(\Omega)F(\Omega)$ 
enters via the zero temperature kernels $K^{\pm}(\omega,\omega')$ and the 
thermal phonon kernels ${\bar K}^{\pm}(\omega,\omega')$ defined by
\begin{eqnarray}
\label{kernel}
K^{\pm}(\omega,\omega')&=&\int\limits_0^{+\infty}d\Omega\alpha^{2}(\Omega)F(\Omega) \\
                       &\times& \left[\frac{1}{\omega'+\omega+\Omega+i0^{+}}\right. \nonumber\\
		  &\pm&   \left. \frac{1}{\omega'-\omega+\Omega-i0^{+}}\right]\>, \nonumber
\end{eqnarray}
\begin{eqnarray}
\label{TPkernel}
{\bar K}^{\pm}(\omega,\omega')&=&\int\limits_0^{+\infty}d\Omega\frac{\alpha^{2}(\Omega)F(\Omega)}{e^{\Omega/T}-1} \\
                              &\times&\left[\frac{1}{\omega'+\omega+\Omega+i0^{+}}\right. \nonumber \\
                             &\pm& \left.  \frac{1}{\omega'-\omega+\Omega-i0^{+}}\right]\>.\nonumber
\end{eqnarray}
These equations can be solved through iterative methods. Alternatively, one can solve a hybrid set of equations, based on 
solutions already converged on the imaginary axis \cite{Mars88}. We have checked our results by using both methods. 

From the solutions of the real-axis Eliashberg equations at a given temperature $T$ below $T_{c}$ one can compute the ratio of 
the NMR relaxation rate in the superconducting state $R_{s}=1/(T_{1}T)$ to its value in the normal state $R_{n}$ assuming that 
the coupling between the nuclear spins and the conduction electrons is via a contact hyperfine interaction
\begin{eqnarray}
\label{T1T}
\frac{R_s}{R_n}&=&2\int\limits_0^{+\infty}d\omega
    \left(-\frac{\partial f}{\partial\omega}\right) \\
    &\times&\left[(\re N(\omega))^2+(\re M(\omega))^2\right] \nonumber
\end{eqnarray}
(for derivation see, for example, \cite{SM}).

The frequency dependent conductivity is given by \cite{Marsiglio2}
\begin{eqnarray}
\label{conductivity}
&\phantom{+}& \sigma(\nu)=\frac{\omega_{p}^{2}}{8\pi\nu}\left\{\int_{0}^{+\infty}d\omega\tanh\frac{\omega}{2T}\right. \\
	    &\times& \frac{1-N(\omega)N(\omega+\nu)-M(\omega)M(\omega+\nu)}{-iE(\omega)-iE(\omega+\nu)} \nonumber \\
           &+& \int_{0}^{+\infty}d\omega\tanh\frac{\omega+\nu}{2T}            \nonumber \\
           &\times&\frac{1-N^{*}(\omega)N^{*}(\omega+\nu)-M^{*}(\omega)M^{*}(\omega+\nu)}{
-iE^{*}(\omega)-iE^{*}(\omega+\nu)}
                                   \nonumber    \\ 
	   &+&\int_{0}^{+\infty}d\omega\left(\tanh\frac{\omega+\nu}{2T}-\tanh\frac{\omega}{2T}\right) \nonumber    \\
           &\times&\frac{1+N^{*}(\omega)N(\omega+\nu)+M^{*}(\omega)M_{n}(\omega+\nu)}{
			   iE^{*}(\omega)-iE(\omega+\nu)} \nonumber    \\
	   &+&\int_{-\nu}^{0}d\omega\tanh\frac{\omega+\nu}{2T}      \nonumber    \\
	   &\times&\left[\frac{1-N^{*}(\omega)N^{*}(\omega+\nu)-M^{*}(\omega)M^{*}(\omega+\nu)}{
		   -iE^{*}(\omega)-iE^{*}(\omega+\nu)} \right.  \nonumber    \\
           &+&\left. \left.\frac{1+N^{*}(\omega)N(\omega+\nu)+M^{*}(\omega)M(\omega+\nu)}{
    iE^{*}(\omega)-iE(\omega+\nu)}
    \right]\right\} \nonumber
\end{eqnarray}
where $\omega_{p}^{2}$ is the square of the plasma frequency. The quasiparticle energy $E(\omega)$  
appearing in the denominators in (\ref{conductivity}) is defined by
\begin{equation}
\label{ENERGY}
E(\omega)=Z(\omega)\sqrt{\omega^{2}-\Delta^{2}(\omega)}\>,
\end{equation}
where $Z(\omega)$ is the total renormalization function which includes
the electron-phonon interaction and impurity scattering \cite{Marsiglio2}.
In (\ref{ENERGY}) the branch of the square root with positive real part is taken.

\section{Results and discussion}

The computed critical temperatures for our model were $T_{c}=$ 5.06 K, 5.99 K and 8.32 K for $\lambda=$ 2.38 \cite{Hiroi07} (Fig. 25 in \cite{Hiroi07}), 3 
and 5 \cite{Hiroi,Terashima}, respectively.

\begin{figure}[ht]
\includegraphics[width=\linewidth]{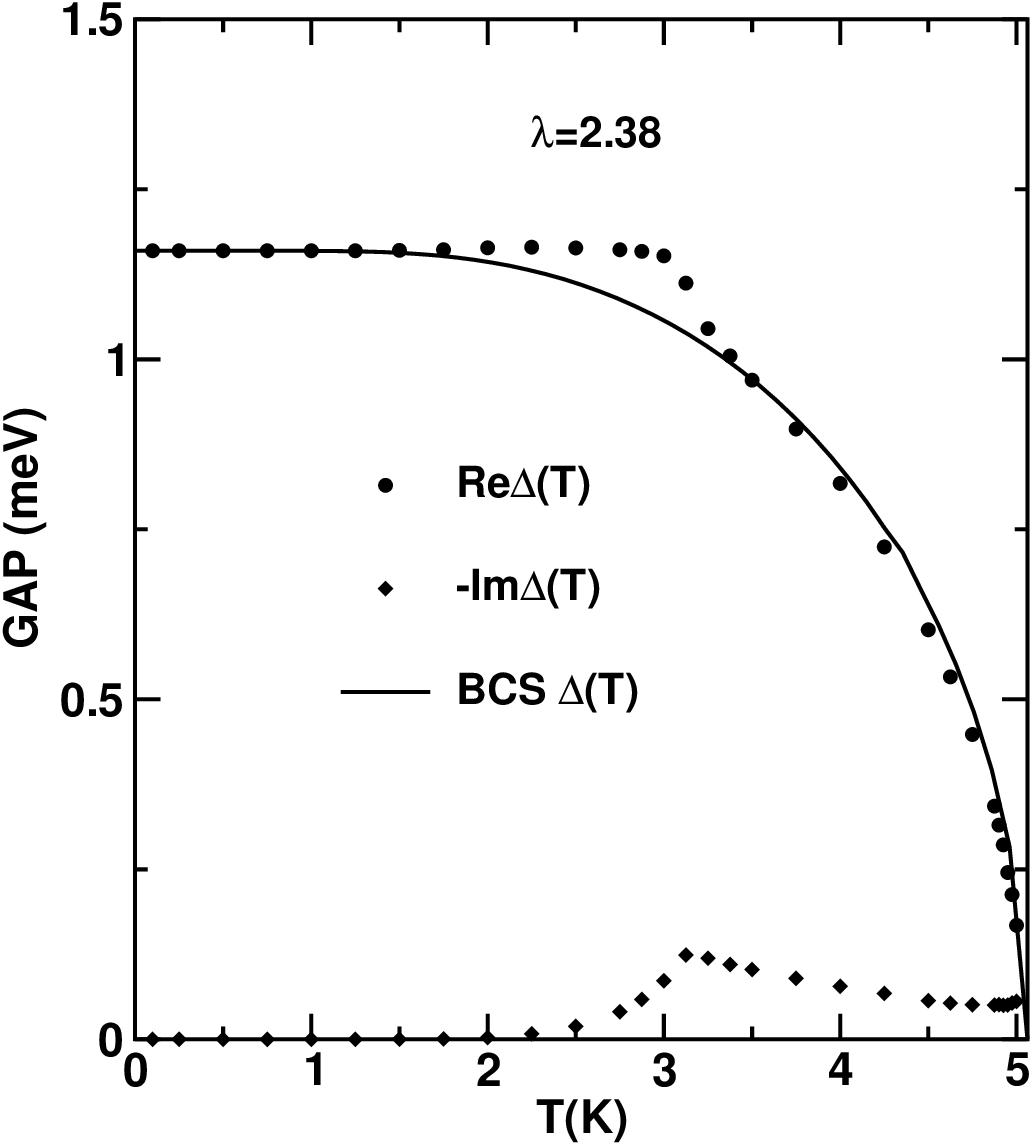}
\topcaption{Temperature dependence of the gap edge $\Delta(T)$ and of the quasiparticle decay rate given by -$\im \Delta(T)$
for $\lambda=$ 2.38. The solid line represents the BCS temperature dependence. }
\label{Fig1}
\end{figure}

\begin{figure}[!ht]
\includegraphics[width=\linewidth]{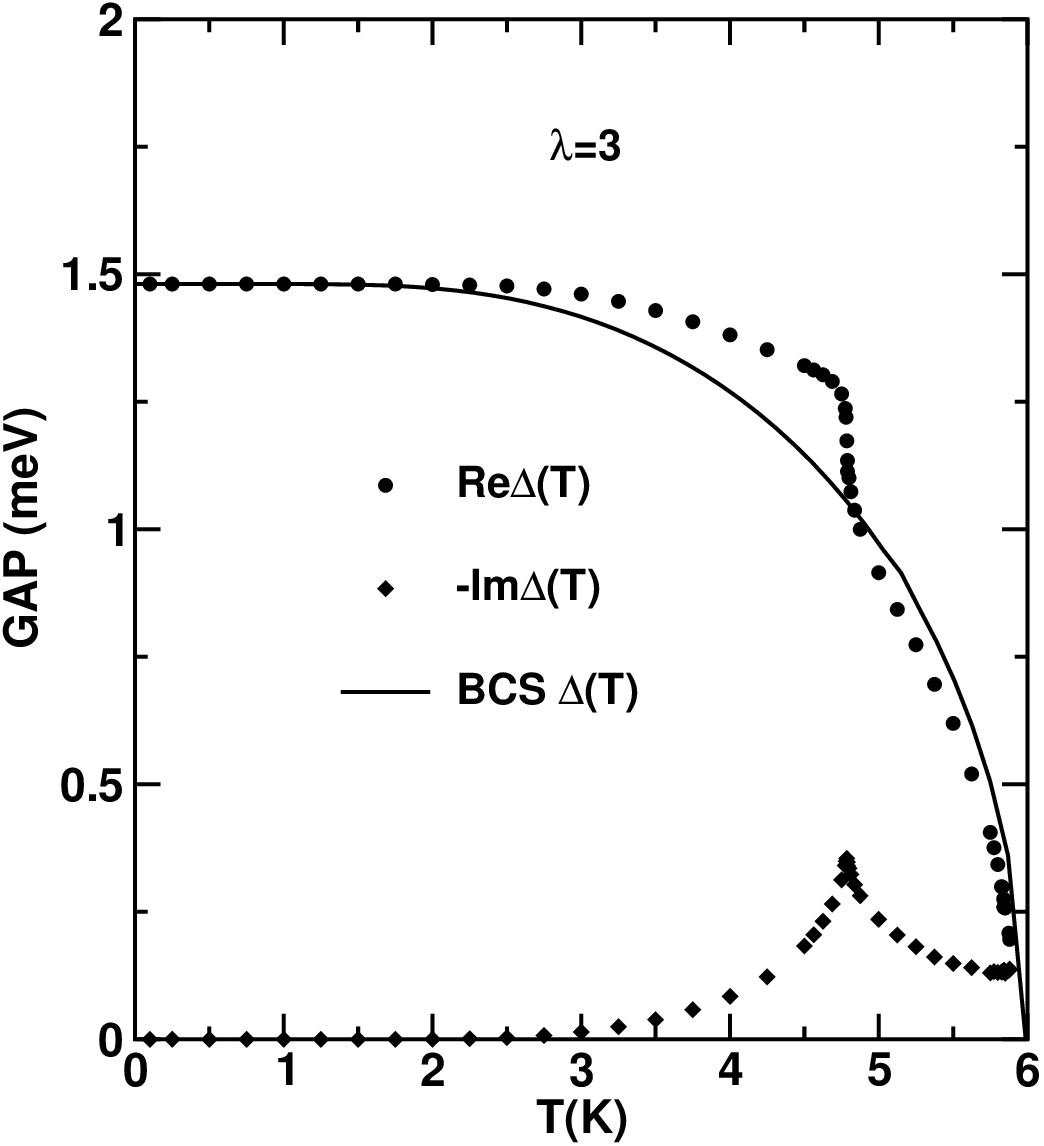}
\topcaption{Temperature dependence of the gap edge $\Delta(T)$ and of the quasiparticle decay rate given by -$\im \Delta(T)$
for $\lambda=$ 3. The solid line represents the BCS temperature dependence. }
\label{Fig2}
\end{figure}

\begin{figure}[h]
\includegraphics[width=\linewidth]{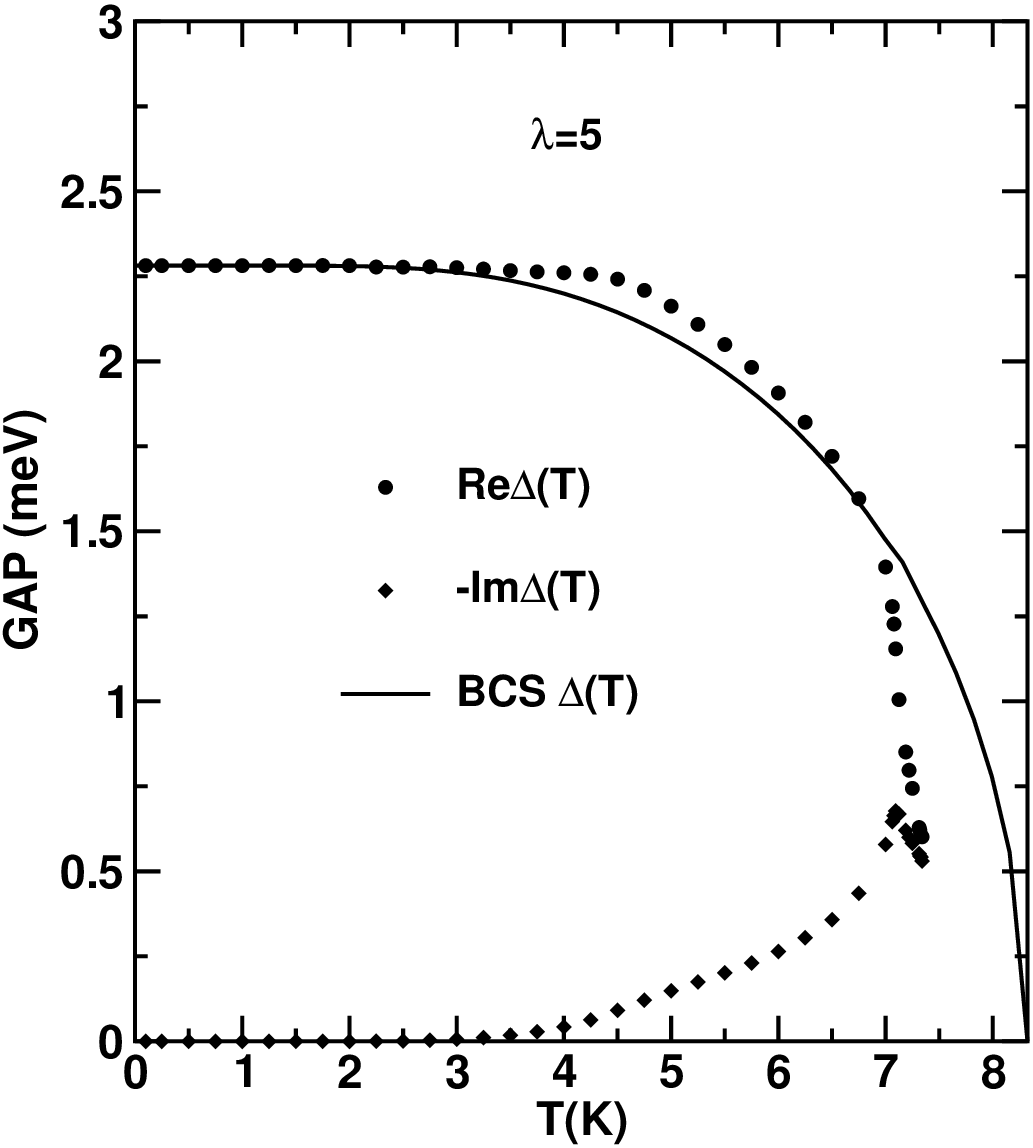}
\topcaption{Temperature dependence of the gap edge $\Delta(T)$ and of the quasiparticle decay rate given by -$\im \Delta(T)$
for $\lambda=$ 5. The solid line represents the BCS temperature dependence. The equation (\ref{gapedge}) did not have a solution for $T>$ 7.34 K.}
\label{Fig3}
\end{figure}

The gap edge at temperature T, $\Delta(T)$ was obtained from solution of
\begin{equation}
\label{gapedge}
\re\Delta(\omega=\Delta(T),T)=\Delta(T)
\end{equation} 
when it existed and the the results are shown in Figs. (\ref{Fig1}-\ref{Fig3}). The computed values for the ratio 2$\Delta(0)/k_{B}T_{c}$ were
5.32, 5.74 and 6.36 for $\lambda=$ 2.38, 3 and 5, respectively, indicating very strong coupling regime in all three cases. 

In Figs. (\ref{Fig1}-\ref{Fig3}) we also show the absolute value of the imaginary part of the gap at the gap edge which rigorously gives 
the quasiparticle decay rate $\Gamma(T)=-\im \Delta(T)$ \cite{Mitrovic}. In all cases considered $\Delta(T)$ deviates from 
the BCS temperature dependence \cite{Muhl}. Note that for very strong coupling and for temperatures very close to $T_{c}$ the density of states 
becomes sufficiently smeared that a `gap' as defined by (\ref{gapedge}) does not even exist \cite{Marsiglio91}.
The non-BCS temperature dependence of the gap shown in Figs. (\ref{Fig1}-\ref{Fig3}) 
and a large damping rate as given by the imaginary part of the gap function 
are quite unusual even for strongly coupled electron-phonon superconductors such as Pb
where $\Delta(T)$ closely follows the BCS curve (see Fig.~44 in \cite{Scalapino}).
A very similar behavior to the one we find was observed in
KOs$_2$O$_6$ ($T_{c}=$ 9.6 K) using photoemission spectroscopy
\cite{Shimojima} (see Fig.~3 in \cite{Shimojima}), and 2$\Delta(0)$/$k_{\rm B}T_c$ for this compound was
estimated to be $\geq$ 4.56. Note that in all three cases shown in Figs. (\ref{Fig1}-\ref{Fig3}) the 
temperature at which the most rapid drop in $\Delta(T)$ sets in coincides with the temperature at which 
the quasiparticle damping rate $\Gamma(T)=-\im \Delta(T)$ attains maximum $\Gamma_{\mbox{max}}(T)$. We 
found $\Gamma_{\mbox{max}}(T)/\Delta(T)$ equal to 11\%, 30\% and 59\% for $\lambda=$ 2.38, 3 and 5, respectively, 
and in the last two cases the quasiparticle picture most definitely breaks down \cite{Kaplan}. In \cite{Shimojima} 
$\Gamma_{\mbox{max}}(T)/\Delta(T)$ for KOs$_2$O$_6$ was found to be $\approx$ 100 \% and the fits to the 
measured density of states using the Dynes formula \cite{Dynes} were clearly invalid (for the validity of the Dynes formula it is 
necessary that $\Gamma(T)/\Delta(T)\ll$ 1 \cite{Mitrovic}). 

\begin{figure}[H]
\includegraphics[width=\linewidth]{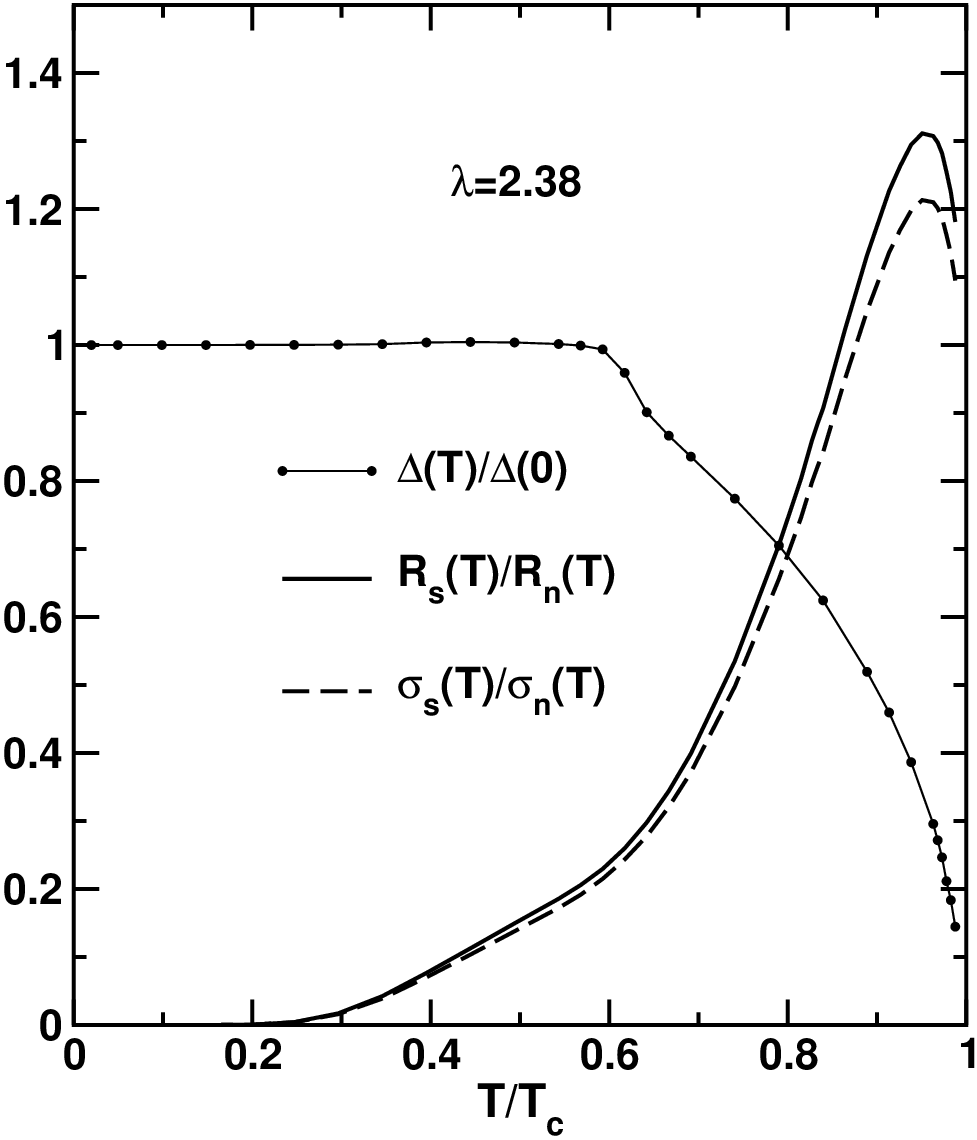}
\topcaption{Numerical results for $\Delta(T)/\Delta(0)$, $R_{s}(T)/R_{n}(T)$, and $\sigma_{s}(T)/\sigma_{n}(T)$ obtained for $\lambda=$ 2.38.}
\label{Fig4}
\end{figure}

\begin{figure}[!b]
\includegraphics[width=\linewidth]{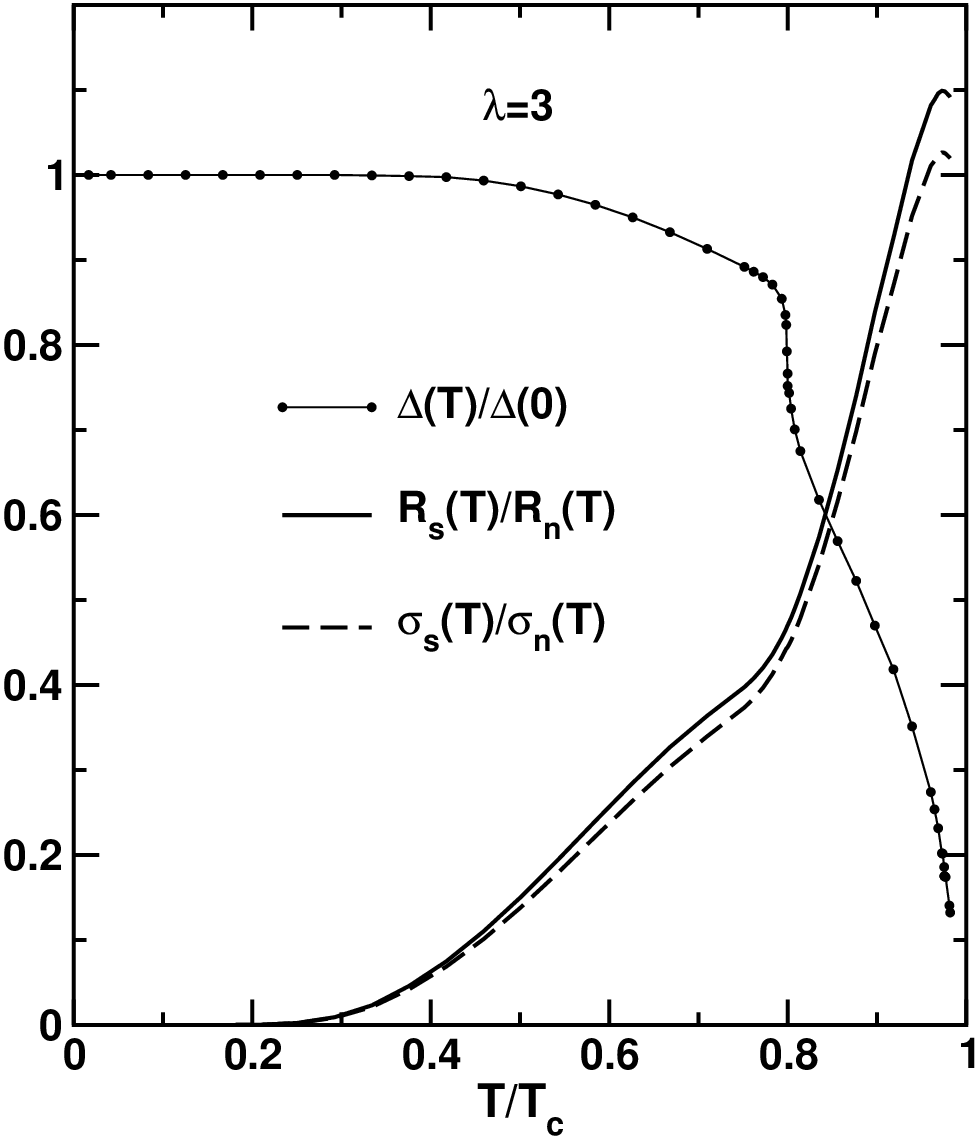}
\topcaption{Numerical results for $\Delta(T)/\Delta(0)$, $R_{s}(T)/R_{n}(T)$, and $\sigma_{s}(T)/\sigma_{n}(T)$ obtained for $\lambda=$ 3.}
\label{Fig5}
\end{figure}

\begin{figure}[H]
\includegraphics[width=\linewidth]{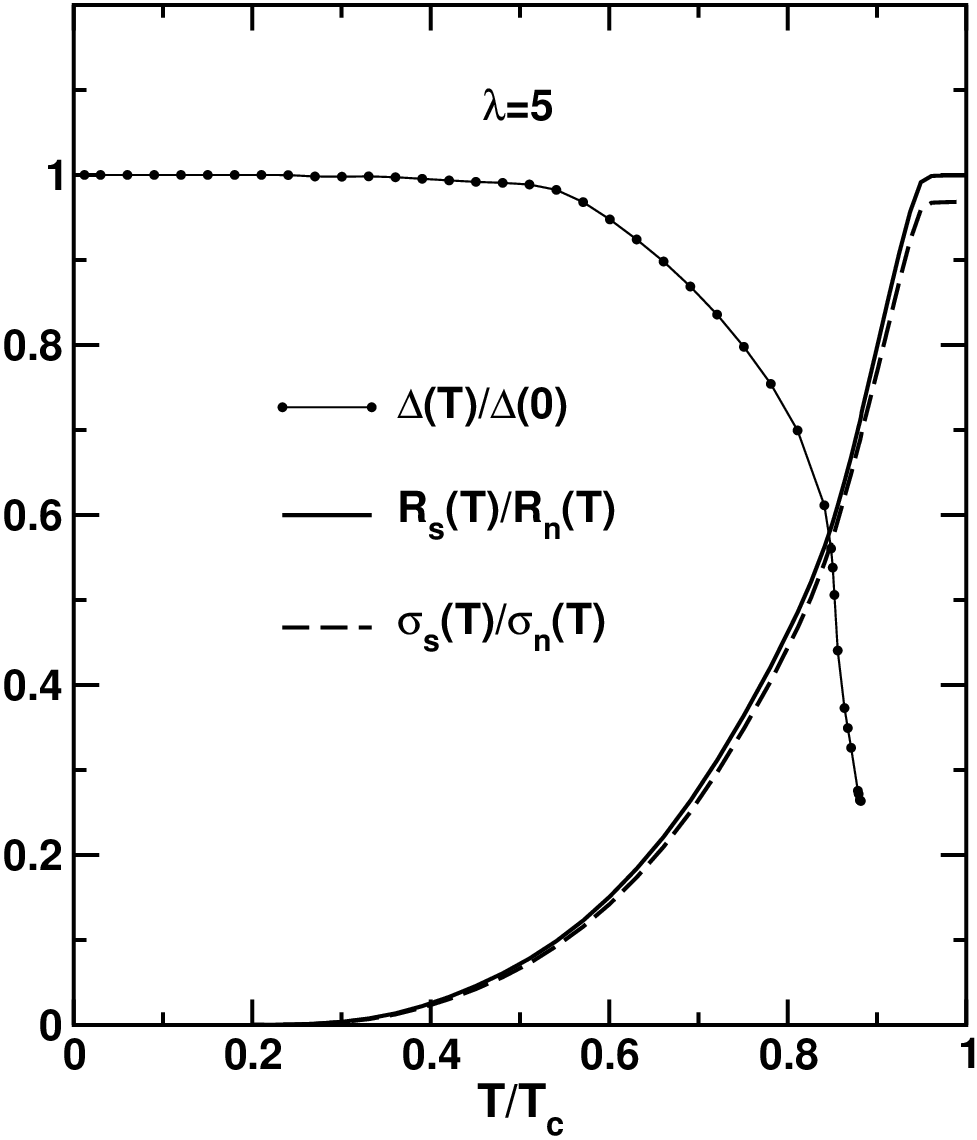}
\topcaption{Numerical results for $\Delta(T)/\Delta(0)$, $R_{s}(T)/R_{n}(T)$, and $\sigma_{s}(T)/\sigma_{n}(T)$ obtained for $\lambda=$ 5.}
\label{Fig6}
\end{figure}

In Figs. (\ref{Fig4}-\ref{Fig6}) we show the NMR relaxation rate $R_{s}(T)$ and the microwave conductivity $\sigma_{s}(T)$, both
normalized to their normal state values, together with $\Delta(T)/\Delta(0)$ as functions of reduced temperature $T/T_{c}$. 
$R_{s}(T)/R_{n}(T)$ and $\sigma_{s}(T)/\sigma_{n}(T)$ have very similar shapes for each value of $\lambda$. The coherence peaks in  
$R_{s}(T)/R_{n}(T)$ and $\sigma_{s}(T)/\sigma_{n}(T)$ at about 95\% of $T_{c}$ are reduced with increasing $\lambda$ and 
disappear for $\lambda=$ 5. 
We note again that the measurements of the NMR relaxation rate on O-sites in KOs$_{2}$Os$_{6}$ \cite{Yoshida}
found that the Hebel-Slichter peak below $T_{c}$ is strongly suppressed, and our results in (\ref{Fig4}-\ref{Fig6}) suggest that for this compound 
$\lambda\geq$ 5 in agreement with specific heat \cite{Hiroi} and Fermi srface \cite{Terashima} measurements.
\newpage

An increase in slope of $R_{s}(T)/R_{n}(T)$ and $\sigma_{s}(T)/\sigma_{n}(T)$
near the temperature where $\Delta(T)/\Delta(0)$ undergoes a sharp drop occurs in all cases, (\ref{Fig4}-\ref{Fig6}), and is most notable for
$\lambda=$ 3.

In conclusion, we have examined the temperature dependence of the superconducting gap edge, the quasiparticle decay rate, the NMR
relaxation rate assuming that the coupling between the nuclear spins and the conduction electrons is via a contact hyperfine interaction, and the
microwave conductivity for a model of electrons coupled only to the rattling phonon. We examined the limit of very strong coupling by
considering three values of electron-phonon coupling parameter $\lambda=$ 2.38, 3, and 5 and {\em did not} assume that the rattler
frequency $\Omega_{0}$ is temperature dependent in the superconducting state. We obtained very unusual temperature dependence of
the superconducting gap edge $\Delta(T)$, very much like the one extracted from photoemission experiments on KOs$_2$O$_6$ \cite{Shimojima}.

{\it Acknowledgments}: Financial support for this work was
partially provided by the Natural Sciences and Engineering Research
Council of Canada.


\begin{thebibliography}{0}

\bibitem{Nagao}
Y. Nagao, J. Yamaura, H. Ogusu, Y. Okamoto, and Z. Hiroi
J. Phys. Soc. Jpn. {\bf 78}, 064702 (2009).

\bibitem{Hiroi}
Z. Hiroi, J. Yamaura, and K. Hattori,
J. Phys. Soc. Jpn. {\bf 81}, 011012 (2012). 

\bibitem{Kunes}
J. Kune\v s, T. Jeong, and W. E. Pickett 
Phys. Rev. B {\bf 70}, 174510 (2004).

\bibitem{Yoshida}
M. Yoshida, K. Arai, R. Kaido, M. Takigawa, S. Yonezawa, Y. Muraoka, and Z. Hiroi,
Phys. Rev. Lett. {\bf 99}, 197002 (2007).

\bibitem{Dahm}
T. Dahm and K. Ueda
Phys. Rev. Lett. {\bf 99}, 187003 (2007).

\bibitem{Shimojima}
T. Shimojima, Y. Shibata, K. Ishizaka, T. Kiss, A. Chainani, T. Yokoya, T. Togashi, 
X.-Y. Wang, C. T. Chen, S. Watanabe, J. Yamaura, S. Yonezawa, Y. Muraoka, Z. Hiroi, 
T. Saitoh, and S. Shin,
Phys. Rev. Lett. {\bf 99}, 117003 (2007).

\bibitem{Scalapino}
D. J. Scalapino, in
{\it Superconductivity},
edited by R. D. Parks
(Marcel Dekker, New York, 1969),
Vol.~1, p.~466.

\bibitem{Grimvall}
G. Grimvall, 
Physica Scripta {\bf 14},63 (1976).

\bibitem{Terashima}
T. Terashima, N. Kurita, A. Kiswandhi, E-S. Choi, J. S. Brooks, K. Sato, J. Yamaura, Z. Hiroi, 
H. Harima, and S. Uji,
Phys. Rev. B {\bf 85}, 180503(R) (2012).

\bibitem{mitrovic84}
B. Mitovi\'c, H. G. Zarate, and  J. P. Carbotte,
Phys. Rev. B {\bf 29}, 184 (1984).


\bibitem{CMM}
J. P. Carbotte, F. Marsiglio, and B. Mitrovi\' c, 
Phys. Rev. B {\bf 33}, 6135 (1986).

\bibitem{Hiroi07}
Z. Hiroi, S. Yonezawa, Y. Nagao, and J. Yamaura,
Phys. Rev. B {\bf 76}, 014523 (2007).

\bibitem{Marsiglio}
F. Marsiglio and J. P. Carbotte,
Phys. Rev. B {\bf 33}, 6141 (1986).

\bibitem{Akis}
R. Akis and J. P. Carbotte, Solid State Commun. {\bf 78}, 393 (1991).

\bibitem{Allen}
P.B. Allen and D. Rainer,
Nature {\bf 349}, 398 (1991).

\bibitem{Mahan}
G. D. Mahan and J. O. Sofo,
Phys. Rev. B {\bf 47}, 8050 (1993).

\bibitem{Marsiglio91}
F. Marsiglio and J. P. Carbotte,
Phys. Rev. B {\bf 43}, 5355 (1991).

\bibitem{AM}
P.B. Allen and B. Mitrovi\' c,
in {\em Solid State Physics}, Vol. 37 edited by H. Ehrenreich, F. Seitz, D. Turnbull (Academic,
New York, 1982), pp. 1–92.

\bibitem{Carbotte}
J. P. Carbotte, Rev. Mod. Phys. {\bf 62}, 1027 (1990).

\bibitem{Mars88}
F. Marsiglio, M. Schossmann, and J. P. Carbotte,
Phys. Rev. B {\bf 37}, 4965 (1988).

\bibitem{SM}
K.V. Samokhin and B. Mitrovi\' c,
Phys. Rev. B {\bf 72}, 134511 (2005).

\bibitem{Marsiglio2}
F. Marsiglio,
Phys.\ Rev. B {\bf 44}, 5373 (1991).

\bibitem{Mitrovic}
B. Mitrovi\' c and  L. A. Rozema,
J. Phys.: Condens. Matter {\bf 20}, 015215 (2008).

\bibitem{Muhl}
B. M\" uhlschlegel,
Z. Phys. {\bf 155}, 313 (1959).

\bibitem{Kaplan}
S. B. Kaplan, C. C. Chi, D. N. Langenberg, J. J. Chang, S. Jafarey, and D. J. Scalapino, 
Phys. Rev. B {\bf 14}, 4854 (1976).

\bibitem{Dynes}
R. C. Dynes, V. Narayanamurti, and J. P. Garno,
Phys. Rev. Lett. {\bf 41}, 1509 (1978).



\end{thebibliography}
\end{document}